\documentclass[a4paper,12pt]{article}

\def\ad   {a^{\dagger}}

\def\bc   {\overline {c}}

\def\bet  {\beta}

\def\HH {\hat H}

\def\HH {\hat H}

{\it}
{\it}
{\it}
{\it}
{\it}
{\it}

\def\bet {\beta}

\def\de {\delta}

\def\om {\omega}

\usepackage{times}
\usepackage{graphicx}
\begin{document}
\title{  Many-body calculations with Deuteron based single-particle bases 
and their associated natural orbits.}
\author{G. Puddu\\
       Dipartimento di Fisica dell'Universita' di Milano,\\
       Via Celoria 16, I-20133 Milano, Italy}
\maketitle
\begin {abstract}
      We use the recently introduced single-particle states obtained from localized
       Deuteron wave-functions as a basis for nuclear many-body calculations.  
      We show that energies can be substantially lowered if the natural orbits
      obtained from this basis are used.
      We use this modified basis for ${}^{10}B$, ${}^{16}O$ and ${}^{24}Mg$ employing
      the bare $NNLO_{opt}$ Nucleon-Nucleon interaction. The lowering of the energies increases
      with the mass. Although in principle natural orbits require a full scale preliminary many-body
      calculation, we found that an approximate preliminary many-body calculation, with a marginal
      increase in the computational cost, is sufficient.
      The use of natural orbits based on an harmonic oscillator basis leads to a
      much smaller lowering of the energies for a comparable computational cost.
\par\noindent
{\bf{Pacs numbers}}: 21.10.Dr,21-60.Cs, 24.10.Cn
\vfill
\eject
\end{abstract}
\section{ Introduction.}
\bigskip
\par
      In ab-initio nuclear structure methods, the most formidable problem is the
      evaluation of properties of nuclei starting from the nucleon-nucleon interaction
      (even more challenging if the NNN interaction is included).
      One of the most fundamental approaches is the No Core Shell Model (NCSM) (refs.[1]-[4]),
      whereby
      the nuclear Hamiltonian is diagonalized using the Lanczos method by constructing
      a basis up to $N_{max}$ many-body excitations above the lowest configuration in an
      harmonic oscillator (h.o.) basis. This approach has been applied to light nuclei due to the 
      explosive increase of the size of the Hilbert space with the number of particles.
      Other approaches like
      the Coupled-Cluster approach (refs.[5]-[8])  scale polynomially with the size of the single-particle
      space and have been used also for medium-mass nuclei. The Similarity Renormalization Group
      (SRG) (refs.[9]-[13]) and especially its In Medium extensions (IM-SRG) have been shown to be particularly
      promising for medium-mass nuclei for constructing valence  space effective Hamiltonians  
      as input to traditional shell model diagonalization techniques, hence leading to
      low-lying spectra (refs.[14]-[19]).
\par
      Most of these approaches use as a single-particle basis harmonic
      oscillator orbits since they allow an exact separation between intrinsic and center
      of mass motion. It has been recognized that for weakly bound systems the asymptotic
      behavior at large distances of the harmonic oscillator single-particle
      wave-functions is not appropriate (e.g. halo nuclei). For weakly bound systems
      the convergence of observables, sensitive to the tail of the single-particle
      wave-functions, is far from optimal even using a large single-particle basis (ref.[20]).
\par
      Recently in order to overcome this limitation of the harmonic oscillator basis,
      the use of other bases have been explored. Most importantly the natural orbits have
      been implemented and a much more satisfactory convergence has been obtained in the
      study of the halo nucleus ${}^6He$ (ref.[21]) . Natural orbits (refs.[22]-[26])
      can be defined as follows. Once a
      spherical single-particle basis of quantum numbers $nljm$ has been selected
      one can perform a preliminary many-body
      calculation to determine a good approximation to the exact ground-state wave-function
      $|\psi>$. One can construct  the one-body density matrix
       $\rho_{n,n'}=<\psi|\sum_m \ad_{n'ljm}a_{nljm}|\psi>$ for each partial
      wave $lj$, $\ad/a$ being the creation/annihilation operators. We then diagonalize 
      the matrix $\rho$. The eigenvectors obtained in this way
      will define a new single-particle basis $\nu,l,j$ called Natural Orbits (NO).
      These orbitals can be used to redo the many-body calculation. In ref. [21] this method showed
      improved convergence properties. Very recently NO basis have been used also for open systems
      (ref.[27]).
\par
      Recently we have introduced a basis which has the desired asymptotic behavior at large distances
      and gives better binding energies (ref.[28]). This basis called Localized Deuteron Basis (LDB),
      has been obtained by diagonalization
      of the $S$-wave of the Deuteron wave function multiplied by a localizing center of mass 
      wave-function.  In this work we use the LDB to construct the corresponding NO basis.
      We are, in this work, primarily interested in increasing binding energies of nuclei
      thereby decreasing the need to work with large single-particle basis, which are the core
      of the computational cost of many-body calculations. As in refs. [21] we do need a preliminary
      many-body calculation, however the increase in the computational cost is minor.
\par
      As a many-body technique  we use the Hybrid-Multi-Determinant (HMD) method (ref.[29]), which expands
      nuclear eigenstates as a linear combination of Slater determinants of the most generic type,
      symmetries being restored with projectors to good many-body quantum numbers. 
      Quasi-Newtonian optimization methods (refs. [30],[31]) are then used to determine the many-body wave function.
      The projectors we commonly use are projectors to good $z$-component of the angular momentum
      and parity ($J_z^{\pi}$). We can use projectors to good angular momentum, but we decided to
      keep the computational cost reasonably low.
      Broadly speaking, our method consists in the following steps.
      First we construct the  Hamiltonian for A nucleons in the LDB basis, and run
      a preliminary many-body calculation. We use a large number of major shells
      and a small number of Slater determinants, typically $15\div 25$,
      (note that a full scale many-body calculation
      needs the construction of at least few hundreds $J_z^{\pi}$ projected Slater determinants).
      We then construct the approximate density matrix and diagonalize it in order to obtain the new
      basis. Next we rewrite the  Hamiltonian in this basis (which we call LDBNO)
      and redo the many-body calculations. 
      
      We find a sizable increase in the binding energies. We considered in this work
      three nuclei, ${}^{10}B$, ${}^{16}O$ and ${}^{24}Mg$. Remarkably, the gain in binding energies
      compared to the harmonic oscillator basis, increases with the mass, at least in the cases we have
      considered. The NN interaction we have used is the "bare" $NNLO_{opt}$ interaction (ref. [32]).
      The outline of this work is as follows. In section 2 we describe the method, in section 3
      we discuss the numerical results. Particular emphasis is placed to the cases of small number
      of major shells, since for large single-particle spaces we expect all bases to give essentially
      the same results. In section 4 we give some concluding remarks.
\bigskip
\section{ Computational method and choice of the single-particle basis.} 
\bigskip
{\it{ 2a. The Localized Deuteron Basis.}}
\bigskip
\par
      Consider the Hamiltonian in the center of mass system for  $A$ particle interacting with 
      a potential $V_{ij}$, 
      $H=\sum_{i<j}H_{ij}=\sum_{i<j}((\vec p_i-\vec p_j)^2/2mA +V_{ij})$, where $p_i$ is the momentum
      of particle $i$. In ref.[28] we took $A=2$,
      although in principle we could consider $A$ as a variational parameter in order to construct an 
      efficient single-particle  basis.
      Let us diagonalize $H_{12}$ and discard the $D$-wave of the ground-state wave-function.
      The $S$-wave depends on the relative momentum of the neutron and proton and
      it is not localized  in coordinate space. We  achieve  localization by multiplying this two-particle
      wave function by a center of mass wave-function in an $S$ state. The full wave function depends on
      the momenta $\vec p_1,\vec p_2$ through $|\vec p_1-\vec p_2|$ and $|\vec p_1+\vec p_2|$. The cosine
      dependence can be expanded in terms of Legendre polynomials which can then be expanded  
      in terms of spherical harmonics of the angular coordinates of particle $1$ and $2$.
      The coefficients of this expansion depend on the momenta $p_1$ and $p_2$,
      These coefficients (for each single-particle angular momentum value $l$) can be diagonalized on a mesh.
      Thus the full two-particle wave-function is  written 
      as a linear combination of products  of spherical single-particle wave-functions $Q_{n,l}(p)Y_{lm}(\hat p)$
      for particle $1$ and $2$. The quantum number $n$ labels the eigenvalues properly reordered so that 
      the largest absolute values of the coefficients in this linear combination correspond to the smallest values of $n$.  
      The aforementioned center of mass wave function must be such that in coordinate space the single-particle 
      radial part decays as  $exp(-\alpha r)$,
      $\alpha$ being a free positive parameter. Its role is to "squeeze" or extend the size of the system.
      The $n,l$ space orbits are augmented with the spin degrees of freedom giving the single-particle basis
      $n,l,j,m$.
      A full discussion of the properties of this single-particle basis, as well as its nodal structure,
      is given in ref. [28].
\par
      The evaluation of the matrix elements of the interaction in this basis needs some discussion.
      In principle we could use the vector brackets formalism of refs.[33]-[36] in order to evaluate
      the matrix elements $<a,b,J|V|c,d,J>$ for the $nn,np,pp$ cases, for the single-particle states
      $a=(n_a,l_a,j_a),b=(n_b,l_b,j_b),..$. This is the optimal method for strong interactions  at large
      relative momentum transfer. In this work we use the $NNLO_{opt}$ interaction which is sufficiently soft
      so that we can expand the above matrix elements in terms of the corresponding ones in an harmonic
      oscillator basis (ref. [37]). Therefore, we first evaluate the matrix elements $<a',b',J|V|c',d',J>$ in a 
      sufficiently large harmonic oscillator basis, then we evaluate  the sums
$$
<a,b,J|V|c,d,J>=\sum_{a'b'c'd'}<a|a'><b|b'><c|c'><d|d'> <a',b',J|V|c',d',J> 
\eqno(1)
$$
      and the overlaps are given by
$$
<a|a'>=\de_{l_a,l_a'} \de_{j_a,j_a'}\int dp Q_{n_a,l_a}(p) P_{n_a',l_a'}(p)
\eqno(2)
$$
      $P_{n_a',l_a'}$ being the harmonic oscillator radial wave functions.
      We considered the harmonic oscillator matrix elements
      in a basis satisfying $2 n_a'+l_a'+2 n_b'+l_b'\leq N_{2max}$ for $N_{2max}=26$. Softer interactions 
      can presumably be dealt with smaller values of $N_{2max}$. We found that binding energies increase
      with increasing values of $N_{2max}$. This is the reason why we had to consider $27$ major h.o. shells.
      Also note that if $N_{2max}$ is sufficiently large the matrix elements of eq.(1) become independent
      of frequency of the h.o. basis because of completeness. Moreover the frequency
      which appears in the
      overlaps of eq.(2) has no relation  with the frequency of the Hamiltonian of the center of mass 
      (see below).  
      The single-particle space spanned by the LDB, to be used in many-body calculations,
      satisfies the restriction
$$
2 n_a+l_a\leq e_{max}
\eqno(3)
$$
      The comparison between many-body calculations, using the LDB and the ones  obtained with the h.o.
      representation is
      meaningful if the h.o. quantum numbers satisfy the same restriction of eq.(3), i.e.
$$
2 n_a'+l_a'\leq e_{max}
\eqno(4)
$$
      As before $n_a',l_a'$ are the h.o. quantum numbers. Note that the large h.o.space is used only to
      evaluate the matrix elemnts of the interaction in the LDB basis.
      The expansion method of eqs.(1)-(2) is widely used, however for "harder" interactions
      the vector brackets formalism is
      presumably the most appropriate one.
      Note however that the necessary number of h.o. shells to be used in the expansion 
      of eqs.(1) and (2) has some relation to the number of h.o. necessary to properly take into
      account the "hardness" of the NN interaction. For  interactions "harder" than NNLO-opt a much larger
      value of $N_{2max}$ is necessary. 
      The  intrinsic kinetic energy matrix elements are evaluated
      directly with the LDB orbits. We always add to the Hamiltonian  the center of mass term
      $\beta(H_{cm}-3\hbar\om/2)$, where $H_{cm}$ is the harmonic oscillator Hamiltonian for
      the center of mass, in order to suppress center of mass excitations. Its frequency, $\hbar\om$
      is not necessarily related to the frequency  used in the aforementioned expansion.
      Before summarizing the HMD variational method, let us assume that we have constructed the full 
      Hamiltonian both in the LDB and h.o. representation for a specified value of $e_{max}$.
      The h.o. frequency in the h.o. representation is selected so as to minimize the ground-state energy
      obtained with few Slater determinants. Afterwards we keep the same value of $\hbar\om$ for
      all values of $e_{max}$,
      although in principle we could redetermine the optimal value of $\hbar\om$ for every value of $e_{max}$.
      Similarly the value of $\alpha$ specifying the LDB is selected such that it minimizes the ground-state
      energy obtained with few Slater determinants for a specified value of $e_{max}$.
\par
      The NO (natural orbits) corresponding to either the h.o. or to the LDB representation are constructed as follows.
      Let us select a sufficiently large value of $max(2n+l)=e_0$. In the calculations discussed in the 
      next section  we took $e_0=7$ (that is, $8$ major shells).
      Let us consider an approximate eigenstate $|\psi>$ obtained as a linear combination of $N_D$ variationally determined
      Slater determinants (typically $N_D=15$ at the most $N_D=25$)
      and let us evaluate the density  matrix 
$$
\rho_{n,n'}= <\psi|\sum_m \ad_{n'ljm} a_{nljm}|\psi>
\eqno(5)
$$
      for all $l,j$ quantum numbers. We actually use the sum of neutron and proton densities.
      We diagonalize  $\rho_{n,n'}$
      and obtain the eigenvectors $v_{n,\nu}(l,j)$ where $\nu$ labels the eigenvalues. For both the h.o. 
      and the LDB representation we construct the new single-particle basis (the same for neutrons and protons)
$$
|\nu,l,j>=\sum_{n=0}^{(e_0-l)/2} v_{n,\nu}(l,j)|n,l,j>
\eqno(6)
$$
      The expansion of eq. (6) is usefull only if $e_{max}=max(2\nu+l)< e_0$. In this
       case the number of NO shells is less than the
      number of shells in the r.h.s. of eq.(6). That is, we compress the information of $e_0+1$ major shells
      into a smaller number of $e_{max}+1$ NO shells. We found that there is no gain in using eq.(6) if
      $e_{max}= e_0$. Differently stated, $e_0$ is simply a measure of the accuracy 
      of the expansion of eq.(6), and  $e_{max}+1$ is the number of major NO shells used in the many-body
      calculations.
      These new bases of eq.(6) are the NO corresponding to the h.o. or the LDB representation depending on the 
      initial basis.
      We can now re-derive the matrix elements of the Hamiltonian in this new basis (both h.o and LDB) using
      the expansion method. We stress that we do not obtain any improvement in the binding energies if
      $e_{max}=e_0$. 
      In the case of $NNLO_{opt}$, we always use $e_0=7$  and, at the most $e_{max}\leq 6$.
      The advantage  of using NO orbits is that we need to perform a partial many-body calculation
      with $e_0+1$ major shells with  a small number of Slater determinants,
      hence with a small additional computational cost.
\bigskip
\par
{\it{ 2b. A brief recap of the variational method.}}
\bigskip
\par
     We start with Hamiltonians of the form  
$$
\HH = {1 \over 2} \sum _{i,j,k,l} H_{ijkl} \ad_i\ad_j a_la_k
\eqno(7)
$$
     where  $i,j,k,l$ are the single-particle quantum numbers $(n_i,l_i,j_i,m_i),...$ for 
     both neutrons and protons and for a specified $e_{max}$. The sum runs from $1$ up to the total
     number of single-particle states $N_s$. The HMD method (ref.[29]) expands
     eigenstates as
$$
|\psi>= \sum_{d=1}^{N_D} P|U_d>
\eqno(8)
$$
     where $P$ is a projector to good quantum numbers, $|U_d> $ is  a generic Slater determinant for $A$
     particles  written as
$$
|U_d>= \bc_1\bc_2...\bc_A|0>
\eqno(9)
$$
     where $|0>$ is the vacuum and the generalized creation operators $\bc_a, (a=1,2,..A)$ are
     of the type
$$
\bc_a=\sum_i U_{ia}\ad_i
\eqno(10)
$$
     The complex numbers $U_{ia}$ are determined using the quasi-Newtonian method of rank-3 described in detail
     in ref. [31] in order to minimize the expectation value of the energy.
     In eq.(8), $N_D$ should be as large as possible.  
     For a given $e_{max}$ we start with a small number of Slater determinants and we progressively
     increase $N_D$ to larger and  larger values.
     Since the computational cost can be large for large $N_D$ (especially for large $e_{max}$),
     we resort to the energy  variance extrapolation method (EVE) which we briefly describe below.
     This method has been introduced in ref.[38] and progressively improved in refs. [39]-[43].
     The basic idea is that if $|\psi>$ is sufficiently close to an exact eigenstate 
     of eigenvalue $E_0$, then
$$
<\psi|\HH|\psi> = E_0+ a( <\psi|\HH^2|\psi>-<\psi|\HH|\psi>^2)
\eqno(11)
$$
     where $a$ is a constant. Hence we have to plot $<\psi|\HH|\psi>$ vs the energy variance and extrapolate to $0$ variance.
     The intercept with the energy axis will give the eigenvalue.
     If  $|\psi>$ is not sufficiently close to an eigenstate, there are correction terms in eq.(11).
     In practice we use the reordering technique developed in ref.[44].
     Briefly, this technique is as follows.
     Let us assume that we have collected $N_D$ Slater determinants in a specified order. 
     We can construct many-body states 
$$
|\psi_N>=\sum_{d=1}^N c_d |U_d>\;\;\;  (N=1,2,..,N_D)
\eqno(12)
$$
     where the complex numbers $c_d$ are obtained by minimizing the energy.
     For each $N$ we can evaluate eq.(11). However the order of the Slater determinants is
     arbitrary. The reordering technique consists in reordering the Slater determinants so that
     eq.(11) applied to eq.(12) gives an EVE plot as linear as possible. We have applied this technique 
     to our calculations.
     In using the EVE reordering technique 
     we have to be certain to be in the linear regime and to have "small" corrections to $<\psi|\HH|\psi>$,
     i.e. $<\psi|\HH|\psi>- E_0$ must be small. Note however that energies are proportional to the number
      of particles and variances to its square. There are some uncertainties in the extrapolation.
     Ideally we would use the full angular momentum and parity projector in eq. (8) in order to decrease
     the energy in the variational calculation as much as possible, however the integration over
     the Euler angles is computationally expensive. Different approximations to the exact eigenstates
     can give different values of $E_0$, and it is difficult to estimate the error on the extrapolation.
     In the cases we will discuss we estimate the uncertainty of $E_0$ to few $\% $ (i.e. few to several MeV's). 
     Before discussing the numerical results, we stress that the method has two parameters that 
     govern the convergence. One is $e_{max}$ which defines the single-particle space. 
     The other is the number of Slater determinants $N_D$ which governs
     the convergence within the Hilbert space  specified  by the single-particle space. The necessary values
     of $N_D$ can be quite large and this is the reason why the EVE technique is essential. 
\bigskip
\section{ Numerical results.}
\bigskip
\par
     In this section we discuss the calculations for the binding energies of ${}^{10}B,\;{}^{16}O$ and
     ${}^{24}Mg$. The experimental data for the binding energies are from ref.[45].
     The ground-state spin of  ${}^{10}B$ has been a problem in ab-initio calculations using chiral
     NN interactions, both the N3LO and $NNLO_{opt}$. In ref. [46] within the NCSM , using the  N3LO interaction
     (ref. [47]) the
     experimental spin ($3^+$) and the binding energy of $64.75 MeV$, within a few tens of KeV's, have
     been reproduced with the essential addition of the NNN 
     interaction. In the case of the $NNLO_{opt}$  (used in all calculations of this work) the lowest $1^+$
      and $3^+$ states are nearly degenerate
     (ref. [32]) in the NCSM, however the obtained binding energy is too low. 
\renewcommand{\baselinestretch}{1}
\begin{table}
   \begin{tabular}{| c | c | c | c | c| c | c | c| }
          \hline
 $ e_{max}$ & $ N_D$  &     E(ho)  & $E_{CM}$  &  E(LDB)  & $E_{CM}$&   E(LDBNO) & $E_{CM}$\\

$ 4 $ & $ 150 $ & $ -40.35 $ & $     $ & $  -43.63 $ & $     $ & $ -46.25$ & $     $\\
$ 4 $ & $ 200 $ & $ -41.33 $ & $ 0.23$ & $  -44.49 $ & $ 0.40$ & $ -47.11$ & $ 0.35$\\
$ 5 $ & $ 150 $ & $ -45.05 $ & $ 0.25$ & $  -46.93 $ & $     $ & $ -49.15$ & $     $\\
$ 5 $ & $ 200 $ & $        $ & $     $ & $  -47.99 $ & $ 0.38$ & $ -50.03$ & $ 0.33$\\
$ 6 $ & $ 150 $ & $ -47.46 $ & $ 0.25$ & $  -48.53 $ & $     $ & $ -49.81$ & $     $\\
$ 6 $ & $ 200 $ & $        $ & $     $ & $  -49.57 $ & $ 0.35$ & $ -50.66$ & $ 0.32$\\
$ 7 $ & $ 150 $ & $ -49.21 $ & $ 0.26$ & $  -49.63 $ & $     $ & $       $ & $     $\\
$ 7 $ & $ 200 $ & $        $ & $     $ & $  -50.54 $ & $ 0.34$ & $       $ & $     $\\
          \hline
\end{tabular}
\caption {Expectation values of the energy for ${}^{10}B$ for  $J_z^{\pi}=1^+$. All energies are in MeV's.
The harmonic oscillator calculation used $\hbar\om=18$, while the $LDB$ value for $\alpha$ is
$2.7 fm^{-1}$ and $\hbar\om=16$. After each column the corresponding values of $< \bet(H_{CM}-3\hbar\om/2)>$ 
 for $\bet=0.5$ are given.
Two different rows give the results for two different numbers of Slater determinants. Some
values are omitted. The LDBNO values for $e_{max}=7$ are essentially the same for LDB with $e_{max}=7$.
See the text for more explanations and discussions.} 
\end{table}
     In table 1 we show the results of our calculation for $J_z^{\pi}=1^+$ using the h.o., the LDB
     and the LDBNO representations.
     The LDBNO has been constructed from an approximate LDB calculation at $e_0=7$ and $N_D=15$.
     Several comments are in order. First the LDB basis produced better energies than the h.o. representation.
     However for large $e_{max}$ the three representations give nearly the same results.
\renewcommand{\baselinestretch}{1}
\begin{table}
   \begin{tabular}{| c | c | c | c | c| c | c | c| }
          \hline
 $ e_{max}$ & $ N_D$  &     E(ho)  & $E_{CM}$  &  E(LDB)  & $E_{CM}$&   E(LDBNO) & $E_{CM}$\\
$ 4 $ & $ 150 $ & $ -39.39 $ & $     $ & $  -42.81 $ & $     $ & $ -45.54$ & $     $\\
$ 4 $ & $ 200 $ & $ -40.07 $ & $ 0.24$ & $  -43.43 $ & $ 0.33$ & $ -46.24$ & $ 0.26$\\
$ 5 $ & $ 150 $ & $ -44/34 $ & $ 0.21$ & $  -46.76 $ & $     $ & $ -48.79$ & $     $\\
$ 5 $ & $ 200 $ & $        $ & $     $ & $  -47.46 $ & $ 0.28$ & $ -49.66$ & $ 0.24$\\
$ 6 $ & $ 150 $ & $ -47.02 $ & $ 0.20$ & $  -48.58 $ & $     $ & $ -49.51$ & $     $\\
$ 6 $ & $ 200 $ & $        $ & $     $ & $  -49.31 $ & $ 0.23$ & $ -50.30$ & $ 0.22$\\
$ 7 $ & $ 150 $ & $ -48.96 $ & $ 0.20$ & $  -49.82 $ & $     $ & $       $ & $     $\\
$ 7 $ & $ 200 $ & $        $ & $     $ & $  -50.36 $ & $ 0.21$ & $       $ & $     $\\
          \hline
\end{tabular}
\caption {Expectation values of the energy for ${}^{10}B$ for  $J_z^{\pi}=3^+$. All energies are in MeV's.
The harmonic oscillator calculation used $\hbar\om=18$, while the $LDB$ value for $\alpha$ is
$2.7 fm^{-1}$ and $\hbar\om=16$. After each column the corresponding values of $< \bet(H_{CM}-3\hbar\om/2)>$ 
are given.
Two different rows give the results for two different numbers of Slater determinants. Some
values are omitted. The LDBNO values for $e_{max}=7$ are essentially the same for LDB with $e_{max}=7$.
See the text for more explanations and discussions.} 
\end{table}
     This is not very surprising since
     for large single-particle spaces we expect on general grounds the results to be compatible with 
     each other. The LDBNO representation for a given $e_{max}$ is almost equivalent to the LDB results
     with $e_{max}+1$, except for $e_{max}=e_0$. We have to construct the LDBNO basis from a much larger
     LDB space, otherwise we get essentially the same results. All LDBNO bases discussed here have been
     obtained from an approximate LDB calculations at $e_0=7$. The most relevant comparison is with the smallest
     space.
     We stress that our goal is not to do a one-to-one comparison between the results obtained with the h.o.
     representation and the corresponding ones obtained with the LDB and their associated natural orbits.
     Our goal is to obtain a single-particle basis that outperforms the h.o. and once this representation
     is identified we mostly work with this better representatiom. If we would redo all calculations with
     both bases, there would be no point in trying to construct an optimal one, since it would double
     the computational cost. This is the reason why
     some entries are missing in the tables.
     In table 1, notice that the gain in energy with the LDBNO compared to the harmonic oscillator is about
     $14\%$. Moreover note that the results obtained with $N_D=200$ are about $1MeV$ lower than the ones 
     obtained with $N_D=150$. This points out that a much larger number of Slater determinants is needed to
     reach satisfactory convergence. The $J_z^{\pi}=1^+$ results obtained with the LDBNO basis
      has also been dealt with the  reordered  EVE method. In figs.1-3 we show the data obtained with the HMD
      method, the EVE extrapolations (linear and quadratic) for the LDBNO, the LDB and the h.o. bases respectively for
      the $1^+$ state of ${}^{10}B$.
\renewcommand{\baselinestretch}{1}
\begin{figure}
\centering
\includegraphics[width=10.0cm,height=10.0cm,angle=0]{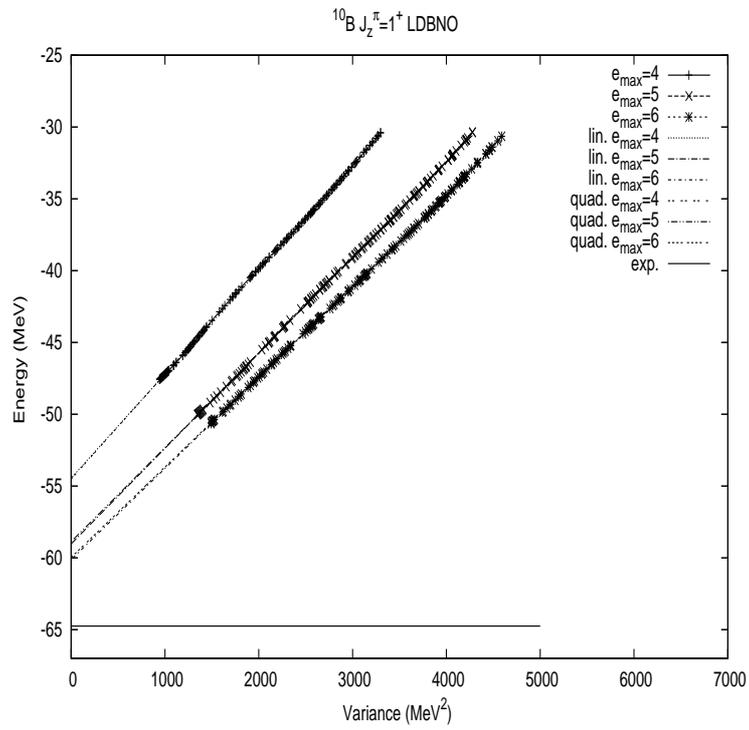}
\caption{Energy vs. variance for ${}^{10}B$ and  $J_z^{\pi}=1^+$ for several $e_{max}$ values. A quadratic fit has also been
 obtained with the LDBNO orbits has also been included.}
\end{figure}
\renewcommand{\baselinestretch}{2}
\renewcommand{\baselinestretch}{1}
\begin{figure}
\centering
\includegraphics[width=10.0cm,height=10.0cm,angle=0]{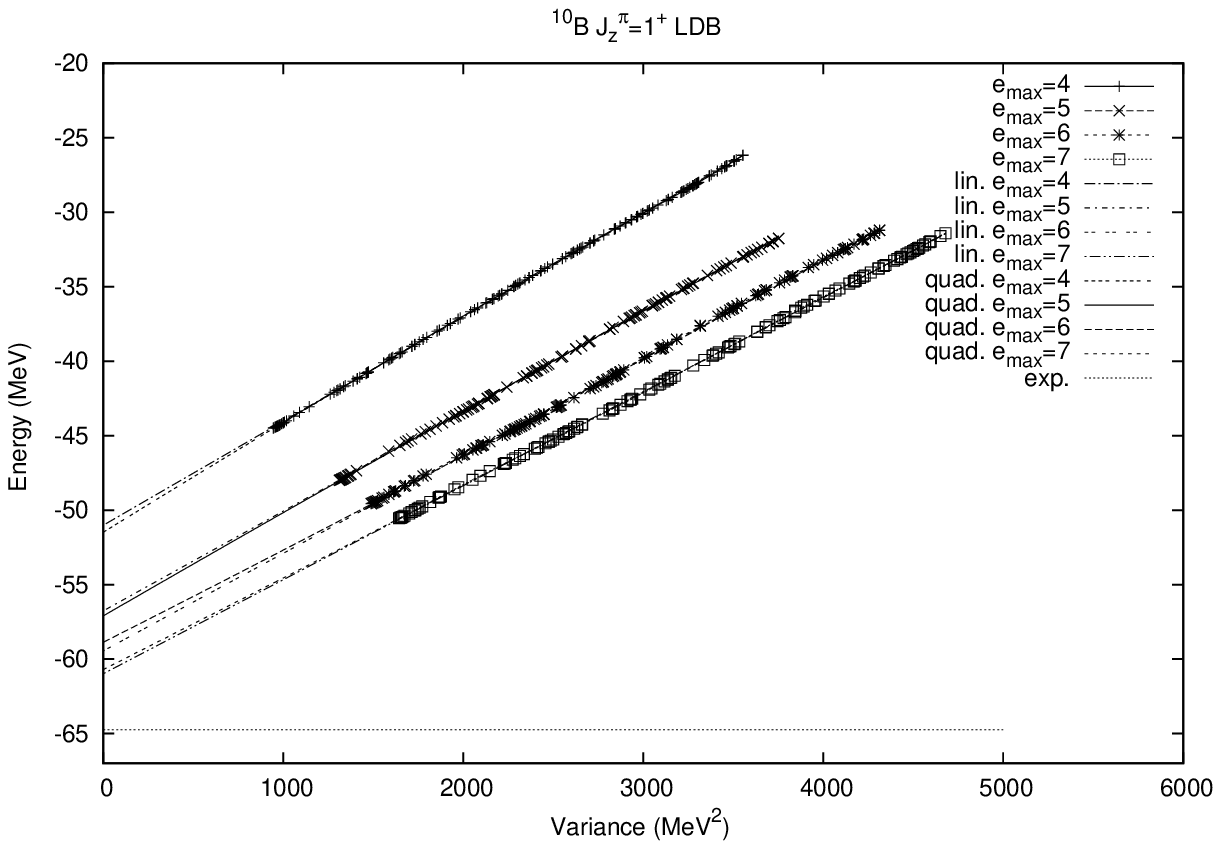}
\caption{Energy vs. variance for ${}^{10}B$ and $ J_z^{\pi}=1^+$ for several $e_{max}$ values. A quadratic fit obtained
 with the LDB orbits has also been included.}
\end{figure}
\renewcommand{\baselinestretch}{2}
\renewcommand{\baselinestretch}{1}
\begin{figure}
\centering
\includegraphics[width=10.0cm,height=10.0cm,angle=0]{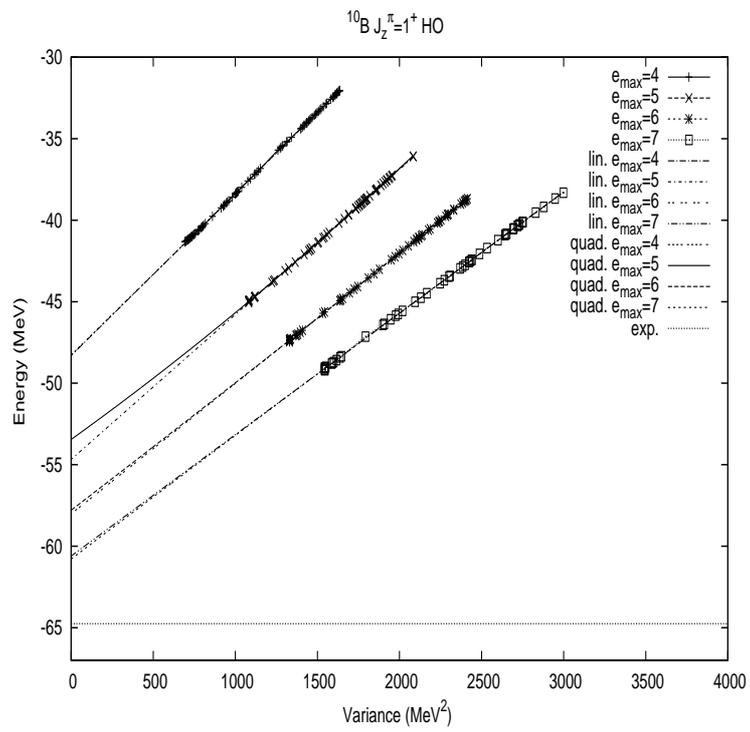}
\caption{Energy vs. variance for ${}^{10}B$ and  $J_z^{\pi}=1^+$ for several $e_{max}$ values. A quadratic fit 
  obtained with the h.o. orbits has also been included.}
\end{figure}
\renewcommand{\baselinestretch}{2}

     As it can be seen from fig. 1, the linear regime 
     is reached especially for $e_{max}=6$. Keeping in mind that the correction to the energy is sizable
     the extrapolated value for the $J_z^{\pi}=1^+$ energy is $-60.1 MeV$, to be compared with  the NCSM 
     results of $-54.35 MeV$ (ref.[32]) obtained for $N_{max}=10$.
     Of course we must keep in mind that the correction given by the EVE
     is quite large and 
     that the presence of additional terms in eq.(11) can modify the extrapolated value. That is, although in principle
     the EVE technique is conceptually very robust, we must be close to $0$ variance.
     The uncertainty can be reduced using the full angular momentum projector. Note that for $e_{max}=4$ there are
     some overtones in fig.1. 
      Note that the cluttering in the figure is only apparent. Consider for example fig. 1 obtained
      for the $1^+$ state of ${}^{10}B$. Fig. 1 consists of three "lines". Each "line" represents the calculations
      for a specified $e_{max}$.  
      The lower the energies of the lines  the larger $e_{max}$. 
      The dots on each line represent the actual computations.
      The upper line is for $e_{max}=4$ and the lower is for $e_{max}=6$ or  $e_{max}=7$, depending on the case.
      Each data line contains 
      two fits: the linear and the quadratic fit. Although it might seem difficult to distiguish 
      the linear from the quadratic fit, showing both of them  gives an idea about the uncertainty of the fit.
      Similar considerations apply to all other figures.  
     In fig. 2 we show the $EVE$ plot in the case of the $LDB$ basis and in fig. 3 the corresponding one for the h.o. basis.
     In all cases of figs. 1-3 for large $e_{max}$ the extrapolated energies are in 
     good agreement with each other, as expected.
     The discrepancies between linear and quadratic fits in the EVE plots give an idea about the uncertainties.
\par
     In table 2, we show the results for the $3^+$ state of ${}^{10}B$
     For the $J_z^{\pi}=3^+$, the EVE plots do not show a linear behavior, in some cases not even monotonic.
     Presumably, the HMD calculations have to be carried out to better accuracy with a much larger number of Slater
     Determinants.
\par
     In table 3 we show the results for ${}^{16}O$.
\renewcommand{\baselinestretch}{1}
\begin{table}
   \begin{tabular}{| c | c | c | c | c | c | c| c |}
          \hline
  
 $e_{max}$ &  $N_D$  &     E(ho)  & $E_{CM}$  &  E(LDB)  & $E_{CM}$&   E(LDBNO) & $E_{CM}$\\
$ 4 $ & $ 150 $ & $ -97.88  $ & $ -   $ & $  -101.89 $ & $     $ & $ -112.82$ & $     $\\
$ 4 $ & $ 200 $ & $ -99.10  $ & $ 0.22$ & $  -102.61 $ & $ 0.51$ & $ -113.84$ & $ 0.25$\\
$ 5 $ & $ 150 $ & $ -105.65 $ & $ 0.27$ & $  -112.17 $ & $     $ & $ -118.00$ & $     $\\
$ 5 $ & $ 200 $ & $         $ & $     $ & $  -113.02 $ & $ 0.31$ & $ -118.65$ & $ 0.19$\\
$ 6 $ & $ 150 $ & $ -108.44 $ & $ 0.26$ & $  -116.75 $ & $     $ & $ -118.97$ & $     $\\
$ 6 $ & $ 200 $ & $         $ & $     $ & $  -117.47 $ & $ 0.21$ & $ -119.70$ & $ 0.16$\\
$ 7 $ & $ 150 $ & $ -110.57 $ & $ 0.27$ & $  -119.20 $ & $     $ & $        $ & $     $\\
$ 7 $ & $ 200 $ & $         $ & $     $ & $  -119.91 $ & $ 0.16$ & $        $ & $     $\\
          \hline
\end{tabular}
\caption {Expectation values of the energy for ${}^{16}O$. All energies are in MeV's.
The harmonic oscillator calculation used $\hbar\om=24$, while the $LDB$ value for $\alpha$ is
$2.5 fm^{-1}$ and $\hbar\om=20$. After each column the corresponding values of $< \bet(H_{CM}-3\hbar\om/2)>$ are
 given.}
\end{table}
       Note that binding obtained for $e_{max}=4$ is about $15\%$ better than the one obtained
       with the h.o. basis.
      In figs. 4-6 we show the EVE plots for ${}^{16}O$ for the LDBNO, the  LDB and the h.o. basis, respectively.
      Again we perform also a quadratic fit to have an idea about the uncertainties. Note for $e_{max}=7$
      the extrapolated ground-state energies are consistent with each other for all bases.
      That is, the $NNLO_{opt}$ interaction is rather soft.
      The extrapolated value for the ground state energy of ${}^{16}O$ is about $-137.6 MeV$. This value should be
      compared with the coupled-cluster value of about $-131 MeV$ obtained with 15 major h.o. shells of ref.[32].  
\par
 In table 4 we present the results for ${}^{24}Mg$. In this case
       we did not perform all calculations with the h.o. basis. For $Mg$ we constructed also a NO basis
       starting from the h.o. basis (under the heading $E(ho-no)$ ) for $e_0=7$ and $N_D=15$ (as for the LDB). Note that the energy
       is higher than the one obtained with the LDB as shown in table 4.
\renewcommand{\baselinestretch}{1}
\begin{table}
   \begin{tabular}{| c | c | c | c | c | c | c| c | c|}
          \hline
 $e_{max}$ &  $N_D$  &     E(ho)  & $E_{CM}$ & E(ho-no)  &  E(LDB)  & $E_{CM}$&   E(LDBNO) & $E_{CM}$\\
$ 4 $ & $ 150 $ & $ -121.58  $ & $ 0.26 $ & $-128.06$ & $ -136.76 $&$      $ & $ -145.58 $ & $     $\\
$ 4 $ & $ 200 $ & $ -122.90  $ & $      $ & $       $ & $ -137.73 $&$0.44  $ & $ -146.84 $ & $ 0.44$\\
$ 5 $ & $ 150 $ & $          $ & $      $ & $       $ & $ -149.26 $&$      $ & $ -156.58 $ & $     $\\
$ 5 $ & $ 200 $ & $          $ & $      $ & $       $ & $ -150.35 $&$ 0.47 $ & $ -157.63 $ & $ 0.53$\\
$ 6 $ & $ 150 $ & $          $ & $      $ & $       $ & $ -157.03 $&$ 0.41 $ & $ -161.02 $ & $ 0.40$\\
$ 7 $ & $ 150 $ & $          $ & $      $ & $       $ & $ -161.04 $&$ 0.38 $ & $         $ & $     $\\
          \hline
\end{tabular}
\caption {Expectation values of the energy for ${}^{24}Mg$. All energies are in MeV's.
The harmonic oscillator calculation used $\hbar\om=20$, while the $LDB$ value for $\alpha$ is
$2.9 fm^{-1}$ and $\hbar\om=16$. After each column the corresponding values of $< \bet(H_{CM}-3\hbar\om/2)>$ are
 given. We included also a result for $5$  natural orbits major shells built from the h.o. $e_{max}=7$ representation
using the expansion method ($E(ho-no)$).}
\end{table}
\renewcommand{\baselinestretch}{1}
\begin{figure}
\centering
\includegraphics[width=10.0cm,height=10.0cm,angle=0]{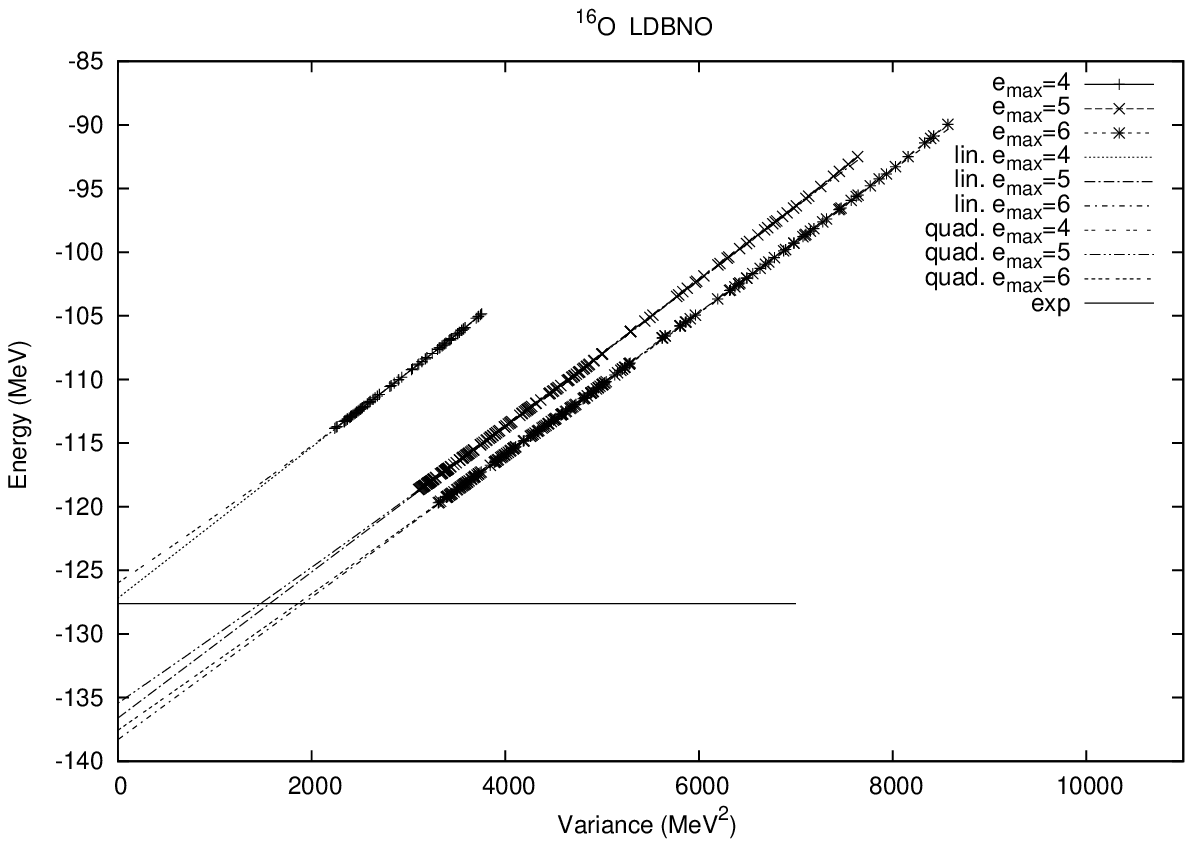}
\caption{Energy vs. variance for ${}^{16}O$ for several $e_{max}$ values. A quadratic fit has also been
 included obtained with the LDBNO orbits.}
\end{figure}
\renewcommand{\baselinestretch}{2}
\renewcommand{\baselinestretch}{1}
\begin{figure}
\centering
\includegraphics[width=10.0cm,height=10.0cm,angle=0]{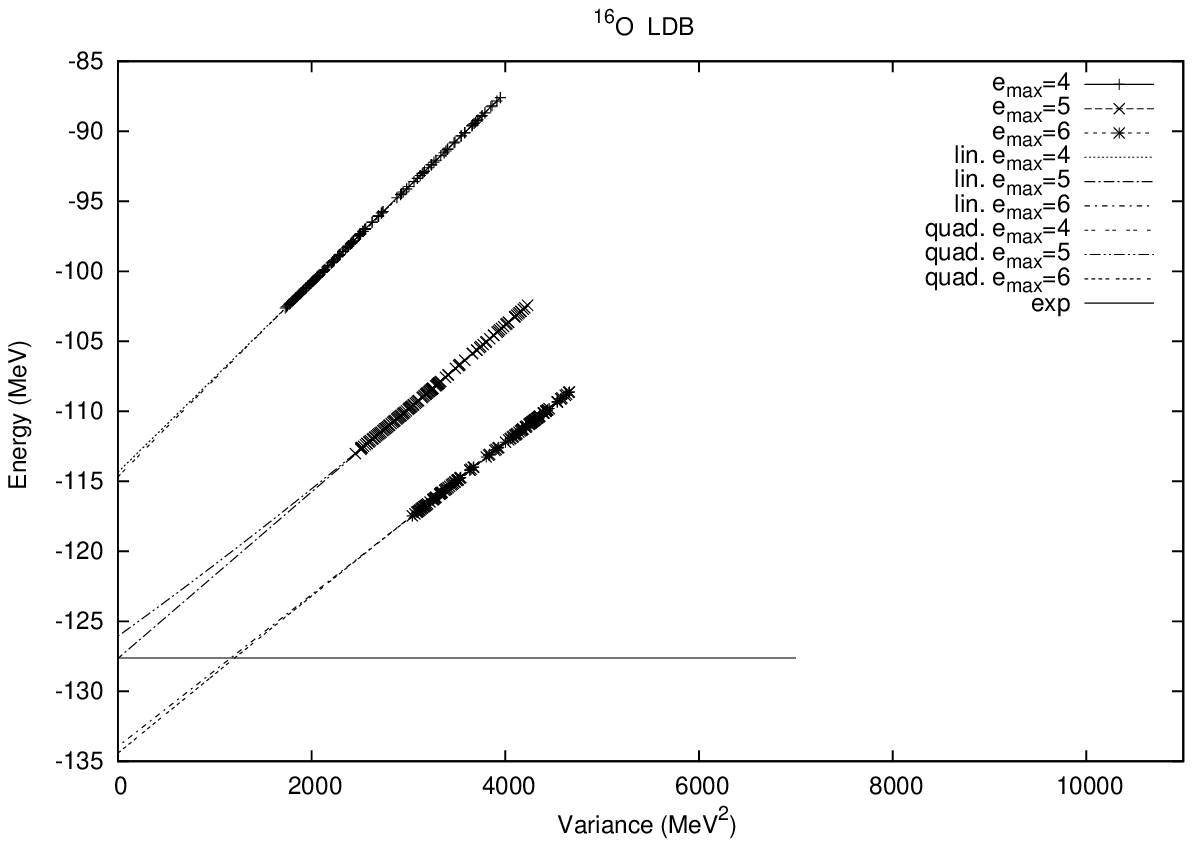}
\caption{Energy vs. variance for ${}^{16}O$ for several $e_{max}$ values. A quadratic fit has also been
 included obtained with the LDB orbits.}
\end{figure}
\renewcommand{\baselinestretch}{2}
\renewcommand{\baselinestretch}{1}
\begin{figure}
\centering
\includegraphics[width=10.0cm,height=10.0cm,angle=0]{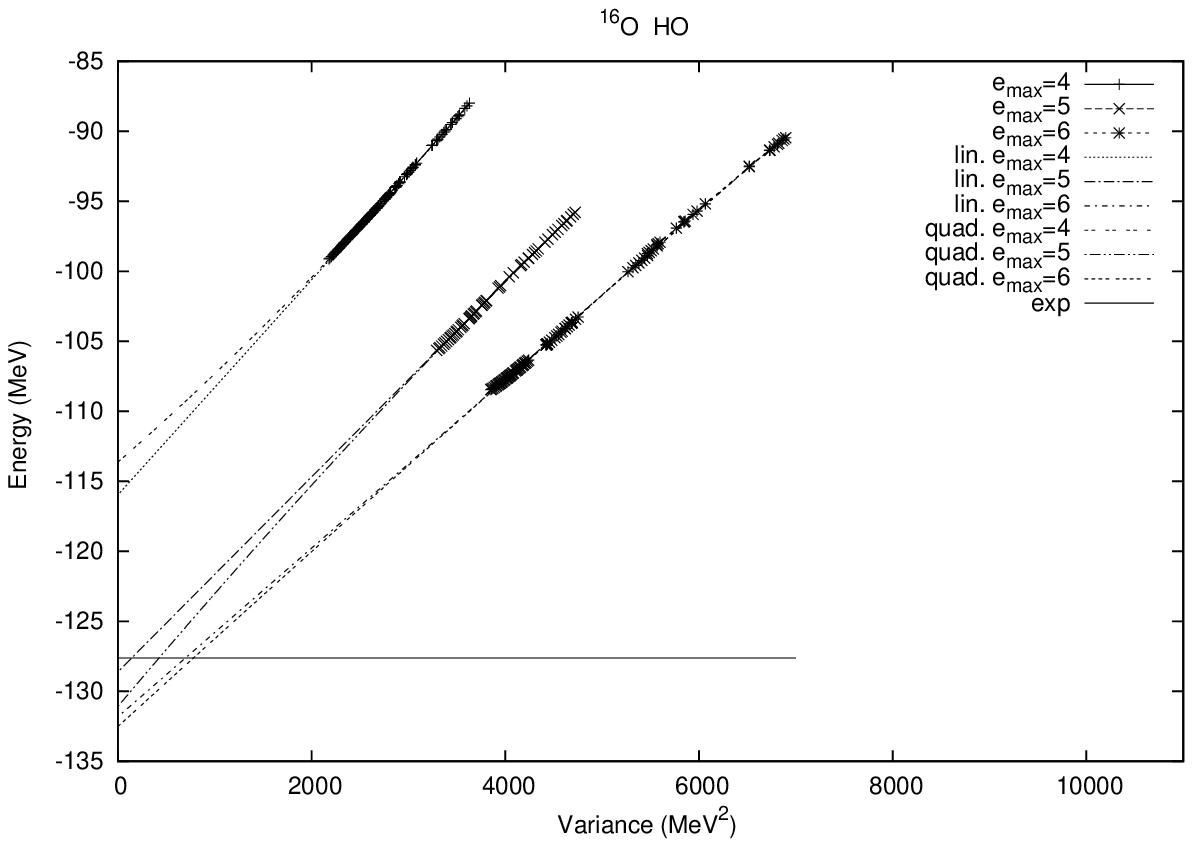}
\caption{Energy vs. variance for ${}^{16}O$ for several $e_{max}$ values. A quadratic fit has also been
 included obtained with the h.o. orbits.}
\end{figure}
\renewcommand{\baselinestretch}{2}
\par
      Note that the increase in binding energy obtained
      with LDBNO is about $19\%$ compared with the h.o. results.
      In  fig. 7 we show the LDBNO EVE plot. Some discrepancies between the linear and the quadratic fit can be seen.
      The extrapolation to $0$ variance for the largest single-particle space, points out to overbinding by
      the $NNLO_{opt}$ interaction
      for ${}^{24}Mg$ with respect to the experimental binding energy.  
\renewcommand{\baselinestretch}{1}
\begin{figure}
\centering
\includegraphics[width=10.0cm,height=10.0cm,angle=0]{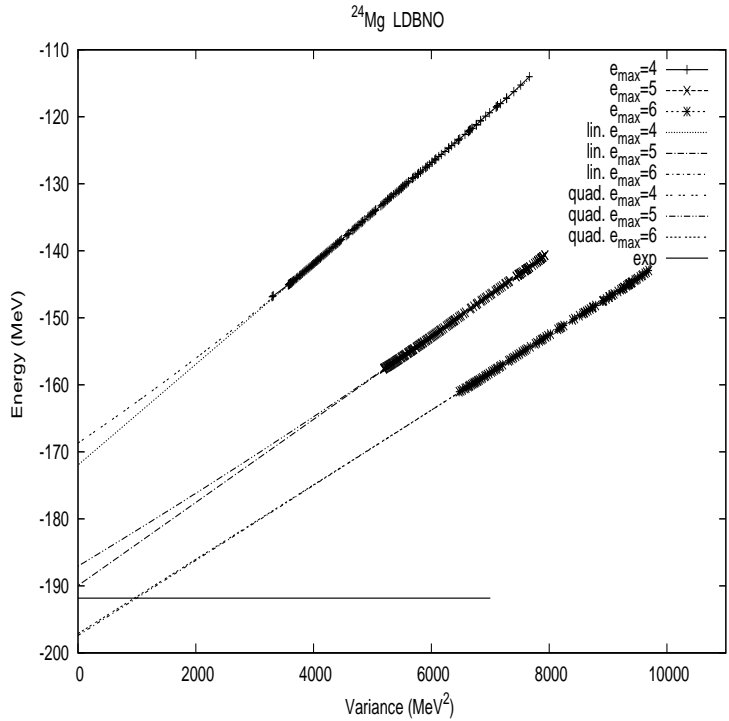}
\caption{Energy vs. variance for ${}^{24}Mg$ for several $e_{max}$ values. A quadratic fit has also been
 included obtained with the LDBNO orbits.}
\end{figure}
\renewcommand{\baselinestretch}{2}
\section {Conclusions.}
      In this work we have performed many-body calculations for three nuclei using a natural orbit single-particle
      basis constructed from the LDB single-particle states. We considered a "bare" NN interaction ($NNLO_{opt}$). A better
      convergence in the binding energies has been obtained. The natural orbits based on the LDB basis outperform
      the standard h.o. representation. The energy gain is more pronounced for the heavier nucleus considered in this work.

\vfill
\eject
\end{document}